\pdfoutput=1

\documentclass{IEEEtran4PSCC}

\ifCLASSINFOpdf
   \usepackage[pdftex]{graphicx}
\else
   \usepackage[dvips]{graphicx}
\fi
\usepackage[cmex10]{amsmath}
\usepackage{amssymb}  
\usepackage{amsfonts}
\usepackage{amsthm}

\usepackage{mathtools} 

\usepackage{graphicx}
\usepackage{epstopdf}
\usepackage{xcolor} 

\usepackage{caption}
\usepackage{subcaption}

\newcommand{\figref}[1]{Fig.~\ref{fig:#1}}
\newcommand{\subfigref}[1]{Fig.~\ref{subfig:#1}}
\newcommand{\secref}[1]{Section~\ref{sec:#1}}

\def \bal#1\eal{\begin{align*}#1\end{align*}}
\def \beql#1#2\eeq{\begin{equation}\label{eq:#1}\begin{aligned}#2\end{aligned}\end{equation}}

\theoremstyle{plain}

\theoremstyle{definition}
\newtheorem{definition}{Definition}
\newtheorem*{problem}{Problem statement}

\theoremstyle{remark}
\newtheorem{remark}{Remark}

\DeclareMathAlphabet      {\mathbfit}{OML}{cmm}{b}{it}

\newcommand{\bsym}{\boldsymbol}
\newcommand{\ubar}{\underline}
\newcommand{\obar}{\overline}

\newcommand{\bjk}{b_{jk}}

\newcommand{\mbR}{\mathbb{R}}

\newcommand{\mcC}{\mathcal{C}}

\newcommand{\mcM}{\mathcal{M}}

\newcommand{\mcX}{\mathcal{X}}

\newcommand{\mcZ}{\mathcal{Z}}

\newcommand{\mxA}{\mathbf{A}}
\newcommand{\mxB}{\mathbf{B}}

\newcommand{\mxF}{\mathbf{F}}

\newcommand{\mxH}{\mathbf{H}}
\newcommand{\mxI}{\mathbf{I}}

\newcommand{\mxM}{\mathbf{M}}

\newcommand{\mxS}{\mathbf{S}}

\newcommand{\slc}{\mathrm{c}}

\newcommand{\slf}{\mathrm{f}}

\newcommand{\slm}{\mathrm{m}}

\newcommand{\slu}{\mathrm{u}}

\newcommand{\vlb}{\mathbfit{b}}

\newcommand{\vlk}{\mathbfit{k}}
\newcommand{\vll}{\mathbfit{l}}

\newcommand{\vlp}{\mathbfit{p}}

\newcommand{\vlu}{\mathbfit{u}}
\newcommand{\vlv}{\mathbfit{v}}
\newcommand{\vlw}{\mathbfit{w}}
\newcommand{\vlx}{\mathbfit{x}}
\newcommand{\vly}{\mathbfit{y}}
\newcommand{\vlz}{\mathbfit{z}}

\newcommand{\vltheta}{\bsym{\theta}}

\newcommand{\vlone}{\bsym{1}}

\newcommand{\suC}{\textsc{C}}

\newcommand{\suF}{\textsc{F}}

\newcommand{\suI}{\textsc{I}}

\newcommand{\suT}{\textsc{T}}

\newcommand{\inv}[1]{#1^{-1}}

\newcommand{\xc}{\vlx_\slc}
\newcommand{\xf}{\vlx_\slf}

\newcommand{\ym}{\vly^\slm}
\newcommand{\yu}{\vly^\slu}
\newcommand{\Mm}{\mxM^\slm}

\newcommand{\Mmc}{\mxM^\slm_\slc}
\newcommand{\Mmf}{\mxM^\slm_\slf}
\newcommand{\Ac}{\mxA_\slc}
\newcommand{\Af}{\mxA_\slf}

\newcommand{\oAu}{\obar{\mxA}_\slu}
\newcommand{\tAm}{\tilde{\mxA}_\slm}

\newcommand{\ymf}{\vly^\slm_\slf}

\newcommand{\txc}{\tilde{\vlx}_\slc}
\newcommand{\txf}{\tilde{\vlx}_\slf}
\newcommand{\tx}{\tilde{\vlx}}

\newcommand{\X}{\bsym{\mcX}}
\newcommand{\Xf}{\bsym{\mcX}_\slf}

\newcommand{\Xc}{\bsym{\mcX}_\slc}

\newcommand{\uc}{\vlu_{\slc}}
\newcommand{\um}{\vlu_{\slm}}
\newcommand{\uu}{\vlu_{\slu}}

\newcommand{\xclb}{\ubar{\vlx}_\slc}
\newcommand{\xcub}{\obar{\vlx}_\slc}

\newcommand{\Ah}{\hat{\mxA}}

\makeatletter
\let\old@ps@headings\ps@headings
\let\old@ps@IEEEtitlepagestyle\ps@IEEEtitlepagestyle
\def\psccfooter#1{%
    \def\ps@headings{%
        \old@ps@headings%
        \def\@oddfoot{\strut\hfill#1\hfill\strut}%
        \def\@evenfoot{\strut\hfill#1\hfill\strut}%
    }%
    \def\ps@IEEEtitlepagestyle{%
        \old@ps@IEEEtitlepagestyle%
        \def\@oddfoot{\strut\hfill#1\hfill\strut}%
        \def\@evenfoot{\strut\hfill#1\hfill\strut}%
    }%
    \ps@headings%
}
\makeatother

\psccfooter{%
        \parbox{\textwidth}{\hrulefill \\ \small{21st Power Systems Computation Conference} \hfill \begin{minipage}{0.2\textwidth}\centering \vspace*{4pt} \includegraphics[scale=0.06]{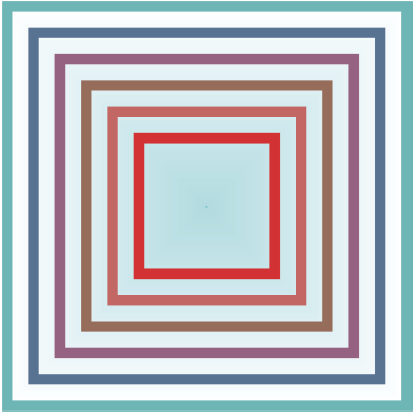}\\\small{PSCC 2020} \end{minipage} \hfill \small{Porto, Portugal --- June 29 -- July 3, 2020}}%
}

\begin{document}
%
\title{On the minimal set of controllers and sensors\\for linear power flow}

\author{
\IEEEauthorblockN{Edwin Mora \& Florian Steinke}
\IEEEauthorblockA{Energy Information Networks and Systems\\
Technische Universit\"at Darmstadt, Germany\\
\{edwin.mora, florian.steinke\}@eins.tu-darmstadt.de}
}


\maketitle

\begin{abstract}
We consider a linear power flow model with interval-bounded nodal power injections and limited line power flows.
We determine the minimal number of power injections to control based on a minimal set of measurements, such that the overall system is feasible for all assignments of the non-controlled power injections.
For the important case where the possible measurements are the nodal power injections, we show that the problem can be solved efficiently as a mixed-integer linear program (MILP).
When also line power flows are considered as potential measurements, we derive an iterative, greedy algorithm that provides a feasible, but potentially conservative solution.
We apply the developed algorithms to both a small microgrid and a modified version of the IEEE 118 bus test power system. 
We show that in both cases a sparse solution in terms of the number of required controllers and measurements can be obtained. 
Moreover, the number of required measurements can be reduced significantly if line flow measurements are considered additionally to nodal power injections.
\end{abstract}

\begin{IEEEkeywords}
Controllability, observability, power flow, resilience
\end{IEEEkeywords}

\thanksto{\noindent This work was sponsored by the German Federal Ministry of Education and Research in project AlgoRes, grant no. 01$|$S18066A.
It has been performed in the context of the LOEWE center emergenCITY.}

\section{Introduction}
Volatile renewable energies are transforming classical power grids with few large generators into complex cyber-physical networks.
These contain a large number of distributed generators and controllable loads, and power lines are often operated close to their limits.
In this context, we ask:
What is the smallest set of generators and/or loads that must be controlled based on the values of a minimal number of measurements, such that (s.t.) the entire system state is feasible for all possible values of the remaining elements?

Being able to identify the (optimally small) set of critical elements in complex power grids reduces the cost and effort for their control.
Moreover, it is an important ingredient to reduce such systems' potentially high vulnerability with respect to (w.r.t.) natural disasters or cyber-attacks \cite{Abedi2019}, enhancing their operational resilience.
An increased protection status could be mandated for the identified critical elements, 
to keep the number of outages and failures in this group at a minimum,
see \cite{Yuan2016} where the hardening of power systems to minimize system damage in case of disasters is examined.

Our research question is an instance of the well-known \emph{optimal input/output selection problem},
also known as the \emph{optimal actuator/sensor placement problem}.
%
Starting with classical work on controllability \cite{Kalman1960}
this problem has attracted long-term research attention, in particular, for linear time-invariant systems.
The problem has recently become very active again in the study of complex networks, see, e.g., \cite{Li2018}.
While most formulations of the problem are NP-Hard due to its combinatorial nature, 
finding only the minimum set of actuators is possible in polynomial-time \cite{Pequito2016}.
This finding is based on structural controllability theory \cite{Lin1974} 
and can be used to develop distributed algorithms for finding the minimum number of controlled and measured nodes \cite{Li2018}.
Structural controllability theory can also be used to analyze cyber-security aspects in distributed power grids \cite{Pasqualetti2015}, e.g., for evaluating the detectability and identifiability of hacked nodes.
Another line of research aims at designing control structures that minimize the control effort, using controllability metrics derived from the controllability Gramian of the system \cite{Lindmark2018}.
Many of the related input/ouput selection problems are submodular
which implies that greedy algorithms using these metrics, e.g., for the optimal placement of High-Voltage direct current lines in a simplified model of the European power transmission network, have provable suboptimality bounds~\cite{Summers2016}.
Time-varying minimal configurations of sensors and actuators can be computed with the help of semi-definite programming \cite{Taha2019}.
All these works are valid for linear (dynamical, algebraic) systems 
without state or input/output restrictions.

In this contribution, we propose an alternative, novel approach based on the steady-state representation of the system only, but considering constrained variables.
This is an important step towards real applications where power injections and line flows are always subject to physical limits.

Our approach extends current work on the distributed control of power systems \cite{Molzahn2017}.
%
For instance, the robust optimal power flow algorithm by \cite{Mesanovic2018} allows computing set points and droop constants for some generators while guaranteeing feasible grid operation for all power injections of other uncertain producers and consumers.
While we use similar modeling, 
we focus on identifying the minimal sets of controllers and measurements that are required for computing such set points.


The rest of the paper is organized as follows.
\secref{lpf} introduces the employed linear power flow model.
The feasibility of a given set of controllers and sensors is defined in \secref{optCM}.
We also give a formal problem statement there as well as further computationally advantageous conditions for testing feasibility.
In \secref{algorithms}, we exploit those conditions for developing two efficient algorithms that minimize the number of controllers and sensors.
In \secref{results}, we apply the proposed algorithms to find the smallest number of controllers and sensors for 1) a simple microgrid consisting of 4 buses and 2) a modified version of the IEEE 118 bus test case.
Finally, concluding remarks and an outlook for future research are provided in \secref{conclusions}.

\section{Linear Power Flow}
\label{sec:lpf}
We analyze an electrical network with $N$ electrical buses connected by $T$ transmission lines
under the common \textit{DC power flow} assumptions \cite{kundur1994power}.
The voltage phase angles $\vltheta\in\mbR^N$ determine the nodal active power injections $\vlp_\suI \in \mbR^N$ and the active power line flows $\vlp_\suF \in \mbR^T$ as
\beql{p2theta}
\vlp_\suI = \mxB_\suI \vltheta, \quad \vlp_\suF = {\mxB}_\suF {\vltheta},
\eeq
where the entries of $\mxB_\suI \in \mbR^{N \times N}$ and $\mxB_\suF \in \mbR^{T \times N}$ are defined element-wise as $B_{\suI,jk} = -b_{jk}$ if $j \neq k$, $B_{\suI,jj} =  \sum_{k} b_{jk}$ and $B_{\suF,jk} = \bjk$, with $\bjk$ the susceptance of the line connecting buses $j$ and $k$.

Without loss of generality, we assume that exactly one generator or load is connected to each bus,
with an externally defined active power set point $x_i$.
If the sum of the set points in the grid is not balanced, a \textit{droop-based primary control} scheme \cite{kundur1994power} adjusts power injections $\vlp_\suI$ under adaptation of the frequency to achieve this balance,
such that in steady state we obtain
\beql{x2p}
\vlp_\suI = \vlx - \vlk \Delta \omega.
\eeq
Here, $\vlk \in \mbR^N$ represents the vector of droop constants, $k_i \geq 0$ and $\sum_i k_i > 0$, and $\Delta \omega \in \mbR$ the frequency deviation with respect to its nominal value.

This common setup implies that the measurable quantities $\vlp_\suI$, $\vlp_\suF$, and $\Delta \omega$ are linearly determined by the controllable quantities $\vlx$.
The kernel of the Laplacian matrix $\mxB_\suI$ contains only the constant vectors for connected graphs, that is, a constant shift of the phase angles has no impact on $\vlp_\suI$. We thus fix $\theta_1 = 0$ and delete the first column of $\mxB_\suI$ to obtain $\tilde{\mxB}_\suI$.
The remaining dimensions of $\vltheta$ are denoted by $\tilde{\vltheta}$. We similarly reduce $\mxB_\suF$ to $\tilde{\mxB}_\suF$.
The image of $\tilde{\mxB}_\suI$ moreover contains all vectors with balanced nodal injections. To handle unbalanced set points $\vlx$, we add $\vlk$ as the last column. This lets us compute for all $\vlx$ with $\cdot$ denoting zero entries
\beql{pL2p}
\begin{bmatrix} \vlp_\suI \\ \vlp_\suF \\ \Delta \omega \end{bmatrix} = 
\begin{bmatrix} \tilde{\mxB}_\suI &  \cdot \\ \tilde{\mxB}_\suF & \cdot  \\ \cdot  & 1 \end{bmatrix} \begin{bmatrix} \tilde{\vltheta} \\ \Delta \omega \end{bmatrix}
= \begin{bmatrix} \tilde{\mxB}_\suI & \cdot \\ \tilde{\mxB}_\suF & \cdot \\ \cdot  & 1 \end{bmatrix} \inv{\begin{bmatrix}\tilde{\mxB}_\suI & \vlk \end{bmatrix}} \vlx.
\eeq
In real systems the nodal injections $\vlp_\suI$ will be limited above and below by the technical capabilities of the connected generator or load. 
Valid set points $\vlx$ might be restricted to smaller intervals than the $\vlp_\suI$, to leave some space for power generation scheduled by the primary controller. 
Similarly, line power flows $\vlp_\suF$ and the frequency deviation $\Delta \omega$ are typically subject to upper and lower bounds.

\section{Feasible Sets of Controllers and Measurements}
\label{sec:optCM}

\subsection{Feasibility Conditions \& Problem Statement}

The power flow model of the previous section can be abstracted as follows:
let $\vlx \in \X \subseteq \mbR^N$ be the variables that can be set externally. 
$\X$ is assumed to be a product of intervals, i.e., $\X = [\ubar{x}_1,\obar{x}_1]  \times \cdots \times [\ubar{x}_N,\obar{x}_N]$.
Variables $\vlx$ can be partitioned into the \emph{controlled} variables $\xc$, for which we will design a controller in the following, and the \emph{free} variables $\xf$, that are left free to be determined either by other users, cooperative or malicious, by fixed external conditions, such as e.g. the weather, or at random.
The index set of the controlled variables is denoted by $\mcC$ and the corresponding partitions of $\X$ as $\Xc$ and $\Xf$.
We assume that the variables $\vlx$ determine the system state uniquely and that the set of feasible system states $\X^*$ can be characterized via a set of linear inequalities, 
\beql{X*}
\X^* = \{ \vlx \in \bsym{\mcX} : \mxA \vlx \leq \vlb \},
\eeq
where $\mxA \in \mbR^{K \times N}$ and $\vlb \in \mbR^K$. 

Similarly, we assume a set of possible measurements $\vly \subseteq \mbR^{L}$ to be linearly related to the system state, i.e., $\vly = \mxM \vlx$ with $\mxM \in \mbR^{L \times N}$.
We partition these possible measurements into the \emph{monitored} measurements $\ym$, that are used as inputs to the control law, and the \emph{unmonitored} variables $\yu$, that are not required for the controller and may or may not be recorded in practice.
The index set of the monitored variables is denoted by $\mcM$.

The defined partitions of $\vlx$ and $\vly$ allow to partition the matrices $\mxA$ and $\mxM$ along their columns or rows as well, yielding $\mxA\vlx = \Ac\xc + \Af\xf$ and $\ym = \Mm \vlx = \Mmc\xc + \Mmf\xf$.

The aim of the paper is to determine the minimal set of controllers $\mcC$ and measurements $\mcM$ that allows for the design of a control law $\xc(\ym)$ that can guarantee a feasible system state, independently of the state of the free variables $\xf$.
This can be formalized as follows.

\begin{definition}[\textbf{Condition} $\suC_1$]
\label{def:A}
Sets $\mcC$ and $\mcM$ are feasible if 
\begin{align}
\exists \xc : \Mmf(\Xf) \rightarrow \Xc \text{ s.t. } \forall \xf \in \Xf: \\
\Ac \xc(\ymf) + \Af \xf \leq \vlb, \nonumber
\end{align}
where $\ymf = \Mmf \xf$.
\end{definition}

The idea behind this definition is that the control $\xc(\ymf)$ chosen for $\ymf$ should be valid for the $\xf$ from which $\ymf$ originated.
Note that we consider only control values for the steady state of the system in this paper. We do not examine whether and how it is possible to get there from arbitrary initial positions. 
Moreover, in order to simplify the notation of the involved sets, we have used only a part of $\ym$ as input to the control law $\xc(\ymf)$. However, since $\ym = \Mmc \xc + \ymf$ one could easily rewrite the controller into the form $\xc(\ym)$, i.e., directly using the measurements that are actually available to the controller.

Since $\xf$ uniquely determines $\ymf$, we can also express the control law as $\xc(\xf)$. 
The formulation $\xc(\ymf)$ implies that $\xc$ will attain the same value for all values of $\xf$ that lead to the same measurements. We thus obtain the following equivalent condition.
\begin{definition}[\textbf{Condition} $\suC_1'$]
\label{def:B}
Sets $\mcC$ and $\mcM$ are feasible if
\begin{align}
\exists \xc : \Xf \rightarrow \Xc \text{ s.t. } \forall \xf,\xf' \in \Xf: \\
\Ac \xc(\xf) + \Af \xf \leq \vlb \; \wedge \nonumber \\
\xc(\xf) = \xc(\xf') \text{ if } \Mmf\xf = \Mmf\xf'. \nonumber
\end{align}
\end{definition}

These definitions  allow us to state the optimization task we aim to solve in this work.
\begin{problem}
Find the set of controllers $\mcC$ and measurements $\mcM$ that solves 
\beql{optA}
            & \min_{\mcC,\mcM}\;{|\mcC| + \gamma |\mcM|} \\
\text{s.t.}\; & \mcC \text{ and } \mcM\text{ are feasible w.r.t. $\suC_1$ or $\suC_1'$}.
\eeq
$|\mcC|$ and $|\mcM|$ denote the cardinality of $\mcC$ and $\mcM$.
The cost of placing a sensor is weighted by $0 \leq \gamma \leq 1$ since it will typically be smaller than implementing a full actuator.

One could additionally incorporate into the objective the varying efforts and costs for controlling certain elements or acquiring certain measurements. Instead of just weighting the total number of controllers and measurements we would then determine an individual weight for each element separately. 
While we do not follow this idea below, all algorithms could straightforwardly be adapted.
\end{problem}

\subsection{Related Conditions}
\label{subsec:relatedCons}

\begin{figure}
\centering
\def\svgwidth{150pt}
\begingroup%
  \makeatletter%
  \providecommand\color[2][]{%
    \errmessage{(Inkscape) Color is used for the text in Inkscape, but the package 'color.sty' is not loaded}%
    \renewcommand\color[2][]{}%
  }%
  \providecommand\transparent[1]{%
    \errmessage{(Inkscape) Transparency is used (non-zero) for the text in Inkscape, but the package 'transparent.sty' is not loaded}%
    \renewcommand\transparent[1]{}%
  }%
  \providecommand\rotatebox[2]{#2}%
  \newcommand*\fsize{\dimexpr\f@size pt\relax}%
  \newcommand*\lineheight[1]{\fontsize{\fsize}{#1\fsize}\selectfont}%
  \ifx\svgwidth\undefined%
    \setlength{\unitlength}{191.21766819bp}%
    \ifx\svgscale\undefined%
      \relax%
    \else%
      \setlength{\unitlength}{\unitlength * \real{\svgscale}}%
    \fi%
  \else%
    \setlength{\unitlength}{\svgwidth}%
  \fi%
  \global\let\svgwidth\undefined%
  \global\let\svgscale\undefined%
  \makeatother%
  \begin{picture}(1,0.13949744)%
    \lineheight{1}%
    \setlength\tabcolsep{0pt}%
    \put(0,0){\includegraphics[width=\unitlength,page=1]{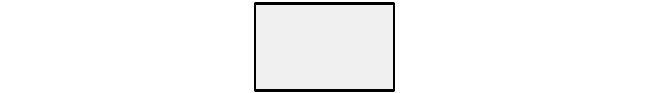}}%
    \put(0.39760187,0.05316082){\color[rgb]{0,0,0}\makebox(0,0)[lt]{\lineheight{1.25}\smash{\begin{tabular}[t]{l}\textit{$\mathrm{C}_1/\mathrm{C}_1'$}\end{tabular}}}}%
    \put(0,0){\includegraphics[width=\unitlength,page=2]{conditions.pdf}}%
    \put(0.83327453,0.05129728){\color[rgb]{0,0,0}\makebox(0,0)[lt]{\lineheight{1.25}\smash{\begin{tabular}[t]{l}\textit{$\mathrm{C}_1^*$}\end{tabular}}}}%
    \put(0,0){\includegraphics[width=\unitlength,page=3]{conditions.pdf}}%
    \put(0.07468499,0.04925742){\color[rgb]{0,0,0}\makebox(0,0)[lt]{\lineheight{1.25}\smash{\begin{tabular}[t]{l}\textit{$\mathrm{C}_2$}\end{tabular}}}}%
    \put(0,0){\includegraphics[width=\unitlength,page=4]{conditions.pdf}}%
  \end{picture}%
\endgroup%

\caption{Relation of the desired conditions $\suC_1$ and $\suC_1'$ to the conditions $\suC_1^*$, $\suC_2$, which are testable with finite resources.}
\label{fig:relation}
\end{figure}

Verifying conditions $\suC_1$ and $\suC_1'$ based on their definition requires checking infinitely many values of $\ymf$ or $\xf$, respectively.
We therefore derive two related conditions that are testable with finite computational resources.
The relation of all derived conditions is presented in \figref{relation}.
In the next section we then show how to exploit them to efficiently solve our problem.

Condition $\suC_1$ requests the existence of a mapping $\xc: \Mmf(\Xf) \rightarrow \Xc$ yielding valid control values. One possibility is that this mapping is affine-linear. 

\begin{definition}[\textbf{Condition} $\suC_2$]
\label{def:C_2}
Condition $\suC_2$ is fulfilled if 
\begin{align}
\exists \mxS \in \mbR^{|\mcC| \times |\mcM|}, \vlw \in \mbR^{|\mcC|} \text{ s.t. }
  \forall \xf \in \Xf: &\\
\xc(\ymf) \in \Xc \; \wedge \; \Ac \xc(\ymf) + \Af \xf &\leq \vlb, \nonumber
\end{align}
where
$\ymf = \Mmf \xf$ and 
$\xc(\ymf) = \mxS \ymf + \vlw$.
\end{definition}

Condition $\suC_2$ is obviously sufficient for $\suC_1$. 
It is, however, not necessary as can be shown by counterexample, 
where piecewise linear control laws sometimes allow for fewer sensors and controllers.
The condition is testable with finite efforts, as we show in \secref{algorithms}.

The conditions presented so far are continuous in the sense that testing their validity requires checking an infinite set of possible realizations of $\xf$ or $\ymf$. However, since the possible values of $\xf$ and $\ymf$ are restricted to bounded polytopes, i.e., $\Xf$ and $\Mmf(\Xf)$,
we can derive a necessary condition for $\suC'_1$ based only on the \emph{corners} of such polytopes.
In contrast to $\suC_2$, such necessary condition will not assume the control law $\uc(\ymf)$ to be affine-linear.

\begin{definition}[\textbf{Corner}]
$\vlz \in \mcZ$ is an \emph{extreme point} or \emph{corner} of the convex set $\mcZ$ if there are no two distinct points $\vlz_1,\vlz_2 \in \mcZ$ and $\lambda \in (0,1)$ such that $\vlz = \lambda \vlz_1 + (1-\lambda)\vlz_2$.
\end{definition}

Denote $C(\Xf)$ as the set containing the corners of $\Xf$. The number of corners of $\Xf$, denoted as $|C(\Xf)|$, is finite, but grows exponentially with the number of free variables.
A condition based on all corners of $\Xf$ would therefore be computationally prohibitive for larger dimensions of $\Xf$. 
Instead, we focus on a subset of corners only, namely those ones which have the maximum impact on the constraints $\mxA \vlx \leq \vlb$.
Denote such subset by $C^A(\Xf)$.
Let $\mxA^i$ be the $i$-th row of $\mxA$, with $i \in \{1,...,K\}$.
Then, a point $\xf$ belongs to $C^A(\Xf)$ if $\xf \in C(\Xf)$ and if there exists $i \in \{1,...,K\}$ such that $\xf$ is an optimal solution for
\beql{maxAfi}
		 \max_{\xf \in C(\Xf)}\;{\Af^i\xf}.
\eeq

\begin{remark}
\label{rem:Bc2Bcrr}
Note that the optimization problem \eqref{eq:maxAfi} defining the elements of $C^A(\Xf)$ can be solved analytically for row $\mxA^i$ as
\beql{XfCA}
x_{j} = \begin{cases} 
      \obar{x}_{j} & A_{ij} \geq 0 \\
      \ubar{x}_{j} & \text{else}
   \end{cases}, j \not\in \mcC.
\eeq
The optimal values thus depend only on the sign of the corresponding elements of $\mxA$. 
In many cases the optimal vectors for different rows of $\mxA$ will therefore coincide and the cardinality of $C^A(\Xf)$ is even smaller than its maximum possible value $K$.
\end{remark}


\begin{definition}[\textbf{Condition} $\suC_1^*$]
\label{def:C_1*}
Condition $\suC_1^*$ is fulfilled if 
\begin{align}
\exists \xc : C^A(\Xf) \rightarrow \Xc \text{ s.t. } \forall \xf,\xf' \in C^A(\Xf): \\
\Ac \xc(\xf) + \Af \xf \leq \vlb \; \wedge \; \nonumber \\
\xc(\xf) = \xc(\xf') \text{ if } \Mmf\xf = \Mmf\xf'. \nonumber
\end{align}
\end{definition}

Conditions $\suC_1/\suC_1'$ straightforwardly imply $\suC_1^*$ since $C^A(\Xf) \subseteq \Xf$.
The reverse is not always true, as can be shown by counterexample. 
However, we will show below that this condition can be exploited for a very efficient computation of approximate sets $\mcC$ and $\mcM$,
at least for the case when the set of possible measurements consists of the power set points at each node,
i.e., when $\mxM$ is an identity matrix of appropriate dimensions, here denoted by $\mxI$.
For nodes with zero droop constant, e.g., typical loads or small-scale generators, the measurement of the power set points is equivalent to measuring nodal power injections.

Minimizing the objective $\mcC + \gamma|\mcM|$ with respect to condition $\suC_1^*$ or $\suC_2$ will provide a lower or upper bound for the optimal solution of problem \eqref{eq:optA}, respectively.
In our experiments we found that for minimal sets $\mcC$ and $\mcM$ fulfilling $\suC_1^*$ it was often possible to determine valid affine-linear control realizations by testing $\text{C}_2$ for such sets, 
i.e., the upper and lower bound coincided.
In this case, $\mcC$ and $\mcM$ are optimal solutions of \eqref{eq:optA}. 

\section{Algorithms}
\label{sec:algorithms}
The feasibility conditions formulated above enable us to develop two methods for addressing optimization task~\eqref{eq:optA}:

The first, derived from condition $\suC_1^*$, leads to a \emph{mixed-integer linear program (MILP)} that finds the smallest feasible sets $\mcC$ and $\mcM$, provided that $\mxM = \mxI$. 
{Since $\suC_1^*$ is necessary for $\suC'_1$ but not sufficient, the obtained sets $\mcC$ and $\mcM$ may be too small to be feasible.}
While we often obtained feasible results anyway, the algorithm can also be used to generate a good initial solution for the second approach.

The second method for solving problem~\eqref{eq:optA} is designed for all possible measurement matrices $\mxM$.
It is a \emph{greedy} procedure based on hill climbing (HC) and condition $\suC_2$.
Recalling that condition $\suC_2$ is sufficient for $\suC_1$ but not necessary, the obtained sets $\mcC$ and $\mcM$ may possibly be too large, but are guaranteed to be feasible. 

\subsection{MILP-Based Approach}
\label{subsec:milpSearch}
In this section, we develop a MILP for finding the smallest feasible sets $\mcC$ and $\mcM$ based on condition $\suC_1^*$, provided that $\mxM = \mxI$.
The key is to formulate condition $\suC_1^*$ as a set of linear inequalities that holds for all choices of sets $\mcC$ and $\mcM$. 

To this end, consider the binary decision variables $\uc \in \{0,1\}^N$ and $\um \in \{0,1\}^N$, defined element-wise as
\begin{equation*}
u_{\slc j} = \begin{cases} 
      1 & j \in \mcC \\
      0 & \text{else}
   \end{cases} , \
u_{\slm j} = \begin{cases} 
      1 & j \in \mcM \\
      0 & \text{else}
   \end{cases},
\end{equation*}
for $j \in \{1,...N\}$.
The decision variables $\uc$ and $\um$ encode the elements of $\mcC$ and $\mcM$, respectively. 
Finding the smallest number of elements of $\mcC$ and $\mcM$ is thus equivalent to minimizing the cost $\|\uc\|_1 + \gamma \|\um\|_1$.

Let $\tx^i \in \X$, $i = 1,\ldots,K$, be defined element-wise as
\begin{equation*}
\tilde{x}^i_j = \begin{cases} 
      \obar{x}_j & A_{ij} \geq 0 \\
      \ubar{x}_j & \text{else}
   \end{cases}.
\end{equation*}
$\tx^i$ is the $\mxA^i$-optimal analytical solution of~\eqref{eq:maxAfi} for the case when all variables are assumed to be free.
Moreover, for any given set of controlled variables $\mcC$, the elements of $C^A(\Xf)$ can be identified with $\txf^i = \tx^i \circ (\vlone - \uc)$, where $\vlone$ is a vector of ones of appropriate dimension and $\circ$ represents the Hadamard product.
Since we assume here that $\mxM = \mxI$, we can further partition the free variables into monitored and unmonitored variables, i.e., we can write $\vlone - \uc = \um + \uu$, with $\uu \in \{0,1\}^N$ being the binary vector that encodes the elements of the unmonitored variables.
The $\mxA^j$-optimal corners of $\Xf$ can then be identified with $\tx^j \circ \uu$, $j=1,\ldots,K$.
Similarly to $\tx^i \circ \um$, a vector of length $N$ whose non-measured entries are zero, we now consider an associated control vector $\txc^i \in \X$ for which
\begin{equation*}
\ubar{\vlx} \circ \uc \leq \txc^i \leq \obar{\vlx} \circ \uc,
\end{equation*}
i.e., $\txc^i$ is a vector of length $N$ whose non-controlled entries are zero.

Condition $\text{C}_1^*$ states that for all $i = 1,\ldots,K$, the control vector $\txc^i$ for the $\mxA^i$-optimal corner $\txf^i$ of $\Xf$ should be valid 
and that it should be the identical to the control vector for all other corners in $C^A(\Xf)$ that cannot be distinguished given the measurements.
We thus can consider only the worst case of the unknown elements and write compactly
\begin{equation*}\begin{aligned}
\mxA \txc^i + \tAm^i \um + \oAu \vlu_\slu \leq \vlb,
\end{aligned}\end{equation*}
with $\tAm^i$ and $\oAu$ defined element-wise as, $k = 1,\ldots,K$,
\begin{equation*}\begin{aligned}
\tilde{A}^i_{\slm , kj} = A_{kj} \tilde{x}^i_j, \quad
\obar{A}_{\slu , kj} = A_{kj} \tilde{x}^k_j.
\end{aligned}\end{equation*}
Thus, the mixed-integer linear program that solves~\eqref{eq:optA} when $\mxM = \mxI$ reads
\beql{MILPMI}
&\min_{\substack{\uc,\um,\uu, \txc^i}} {\|\uc\|_1 + \gamma \|\um\|_1} \\
\text{s.t. } & \mxA \txc^i + \tAm^i \um + \oAu \uu \leq \vlb, & \forall i = 1,...,K,\\
						  & \ubar{\vlx} \circ \uc \leq \txc^i \leq \obar{\vlx} \circ \uc, & \forall i = 1,...,K,\\
							& \uc+\um+\uu = \vlone.
\eeq

\begin{remark}
\label{reductionA}
Note that, particularly in large scale applications, there may be several constraints, i.e., rows of $\mxA$ and corresponding entries of $\vlb$, that are not violated for any realization of $\vlx$. Hence, when optimizing~\eqref{eq:MILPMI}, we only take into account the rows of $\mxA$, for which a violation of~\eqref{eq:X*} is possible, i.e., where $\mxA^i \tx^i - \vlb^i > 0$.
This preprocessing is also utilized by the greedy search proposed below.
\end{remark}

\subsection{Greedy Approach}
\label{subsec:hill}
In this section, we first show how to check condition $\suC_2$ efficiently via a linear program (LP) for fixed sets $\mcC$ and $\mcM$. Thereafter we describe an iterative algorithm to choose and adapt these sets in order to find minimal feasible sets.

For given $\mcC$ and $\mcM$, condition $\suC_2$ mandates to check if there exists a valid affine-linear control law that makes the system feasible for every possible value $\xf \in \Xf$. 
More precisely, there should exist an affine-linear control law defined via $\mxS$ and $\vlw$ such that for all $\xf\in \Xf$ we have
\beql{SM1}
\underbrace{
\begin{bmatrix}
\Ac\mxS\Mmf + \Af \\
\mxS\Mmf\\
-\mxS\Mmf
\end{bmatrix}}_{\Ah(\mxS)} \xf + 
\underbrace{\begin{bmatrix}
\Ac \\
\mxI \\
-\mxI
\end{bmatrix}}_{\mxF} \vlw - 
\underbrace{\begin{bmatrix}
\vlb \\
\xcub \\
-\xclb
\end{bmatrix}}_{\vll} \leq \eta
\underbrace{\begin{bmatrix}
\vlone \\
\cdot \\
\cdot \\
\end{bmatrix}}_{\vlv},
\eeq
where we introduce $\eta \in \mbR$ as an indicator of how far the system is from being infeasible.
A control law is valid if $\eta \leq 0$.

To tackle condition~\eqref{eq:SM1} for all $\xf\in\Xf$ we only need to consider the maximum of the left hand side expression.
Let $\hat{K} = K + 2|\mcC|$ be the number of rows of $\Ah(\mxS)$ and $N_\slf = N - |\mcC|$ the number of free variables.
We can introduce an upper bound on $\Ah(\mxS)\xf$ via a matrix $\mxH \in \mbR^{\hat{K} \times N_\slf}$, whose entries fulfill
\beql{SM5}
H_{ij} \geq \hat{A}_{ij}(\mxS) \obar{x}_{\slf j},  \\
H_{ij} \geq \hat{A}_{ij}(\mxS) \ubar{x}_{\slf j},
\eeq
for all $i = 1,...,\hat{K}$ and $j = 1,...,N_\slf$.
The upper bound of $\Ah(\mxS)\xf$ is then given by $\mxH\vlone$
and condition \eqref{eq:SM1} is equivalent to 
\beql{SM4}
\mxH \vlone + \mxF \vlw - \vll \leq \eta \vlv.
\eeq
Putting these results together allows us to compute the minimum possible value of $\eta$ for given $\mcM$ and $\mcC$ via the following linear program
\beql{optTheta}
& \min_{\substack{\eta, \mxH, \vlw, \mxS}} \; \eta \\
\text{s.t. } & \mxH \vlone + \mxF \vlw - \vll \leq \eta \vlv,\\
& H_{ij} \geq \hat{A}_{ij}(\mxS) \obar{x}_{\slf j}, \forall i = 1,...,\hat{K}, \forall j = 1,...,N_\slf,\\
& H_{ij} \geq \hat{A}_{ij}(\mxS) \ubar{x}_{\slf j}, \forall i = 1,...,\hat{K}, \forall j = 1,...,N_\slf.
\eeq


The above described algorithm for testing the validity of $\suC_2$ for fixed $\mcC$ and $\mcM$ can now be used as a subroutine to minimize over the sets $\mcC$ and $\mcM$ as well. 
To do this, we proceed iteratively from initial sets $\mcC$ and $\mcM$ adapting them one element at a time.
Since we want to measure the optimization progress also for non-feasible combinations $\mcC$ and $\mcM$, we extend the minimization objective to
\beql{costHC}
J(\mcC,\mcM) = |\mcC| + \gamma|\mcM| + \mu \max(\eta, 0),
\eeq
where $\eta$ is the feasibility indicator obtained from solving problem~\eqref{eq:optTheta}.
$\mu >0 $ is a weighting factor that penalizes the infeasibility of $\mcC$ and $\mcM$.
We choose $\mu \gg 1$ to steer the iteration quickly towards feasible solutions.

The cost function~\eqref{eq:costHC} is minimized via a greedy hill climbing procedure.
In each iteration we compute the objective value for all sets $\mcM'$ or $\mcC'$ that can be generated by adding one element to either $\mcM$ or $\mcC$.
We then choose the step which yields the largest improvement of the objective value~\eqref{eq:costHC}.
As soon as the sets of controllers and measurements are feasible, we stop the iteration.

It is well known that the solution of this greedy approach depends on the selection of the starting point.
A natural option is to start with empty sets, selecting the most important controllers and measurements during the first iterations.
Alternatively, we propose to use the MILP~\eqref{eq:MILPMI} formulation as an initial guess.
More specifically, we solve the MILP~\eqref{eq:MILPMI} for $\mxM = \mxI$ first.
We then use the found controller set $\mcC$ as a starting point for the greedy approach, while disregarding the found measurements.
Instead, we start with an empty $\mcM$. 
This way the measurements resulting from general $\mxM$, which potentially allow for more compact control systems than the identity measurements, can be integrated well, but the critical controllers are already identified.

\begin{remark}
Since general MILP has exponential worst-case time complexity, this is an upper bound on the complexity of our first approach \eqref{eq:MILPMI}. 
In contrast, LP as used for our second approach \eqref{eq:optTheta} is known to have polynomial worst-case time complexity,
and the hill climbing procedure only adds polynomial factors.
However, for the realistic examples discussed in the next section we found the MILP approach to be more efficient than the hill climbing procedure. 
The latter's computation time depends strongly on the starting point.
For the examined medium to large problem instances, it allowed finding small, guaranteed to be feasible solutions for $\mcM$ and $\mcC$ with very reasonable efforts.
For cases when $\mxM = \mxI$ the MILP solution could often be verified to be feasible (and thus also optimal) by solving the small LP~\eqref{eq:optTheta} only once without further adaptation of $\mcM$ or $\mcC$. 
We thus see both algorithms as an important contribution for solving real control design problems with state constraints.
\end{remark}
\section{Numerical Examples}
\label{sec:results}
The algorithms developed in~\ref{subsec:milpSearch} and~\ref{subsec:hill} are now applied to find the minimal feasible configuration of controllers and measurements for two exemplary power systems. 
We first demonstrate our setup and typical effects on a simple microgrid of 4 buses connected in a line. Subsequently, a modified version of the IEEE 118 bus test case is addressed.
The experiments were performed using an i5 notebook with 8 GB of RAM.
The algorithms were implemented in Matlab R2018b, using YALMIP \cite{Lofberg2004} as modeling language and CPLEX 12.9 as LP and MILP solver.

\subsection{Simple Microgrid}
\label{subsec:simpleGrid}
\begin{figure*}[h]
\centering
%
\begin{subfigure}[h]{\textwidth}
\centering
\includegraphics[width=0.5\columnwidth]{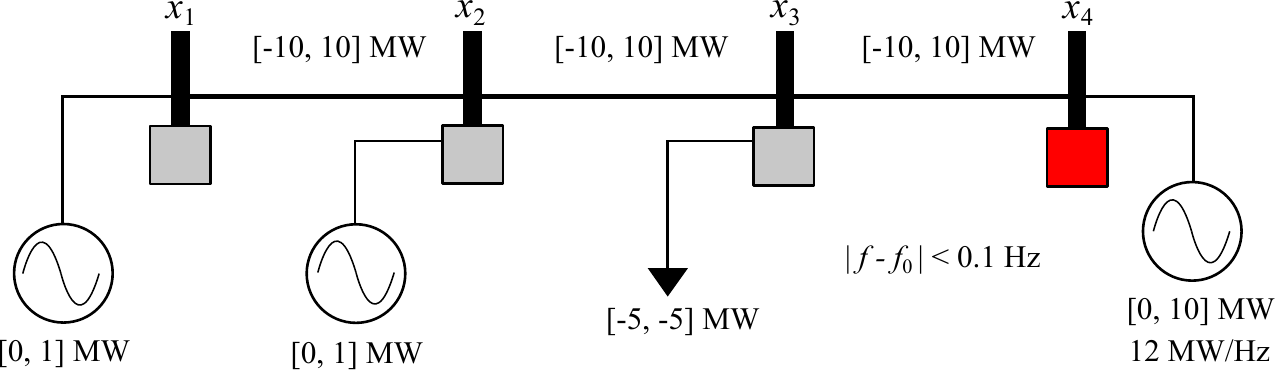} \hspace{3mm}
\includegraphics[width=0.23\columnwidth]{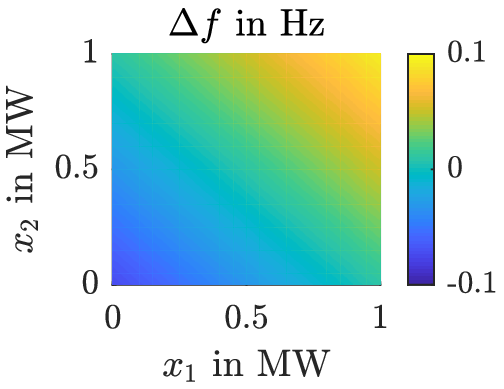}
\subcaption{
$\xc = x_4; \;
\xf =
\begin{bmatrix}
x_1 \! & x_2 \! & x_3
\end{bmatrix}^\suT; \;
\mxS = \emptyset; \;
\Mmf = \emptyset; \;
\vlw = 4.1.$
}
\vspace{3mm}
\label{subfig:simpleGrid1}
\end{subfigure} \\
%
\begin{subfigure}[h]{\textwidth}
\centering
\includegraphics[width=0.5\columnwidth]{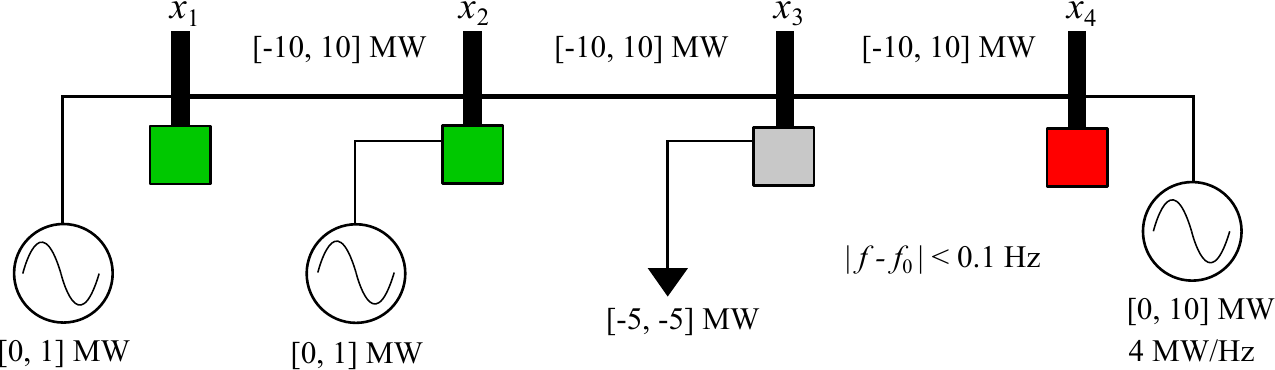} \hspace{3mm}
\includegraphics[width=0.23\columnwidth]{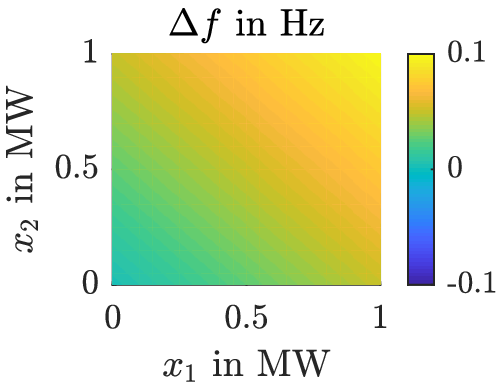}
\subcaption{
$\xc = x_4; \;
\xf =
\begin{bmatrix}
x_1 \! & x_2 \! & x_3
\end{bmatrix}^\suT; \;
\mxS = - 0.81
\begin{bmatrix}
1 & \! 1
\end{bmatrix}; \;
\Mmf = 
\begin{bmatrix}
1 \!& 0 \!& 0\\
0 \!& 1 \!& 0
\end{bmatrix}; \;
\vlw = 5.$
}
\vspace{3mm}
\label{subfig:simpleGrid2}
\end{subfigure} \\
%
\begin{subfigure}[h]{\textwidth}
\centering
\includegraphics[width=0.5\columnwidth]{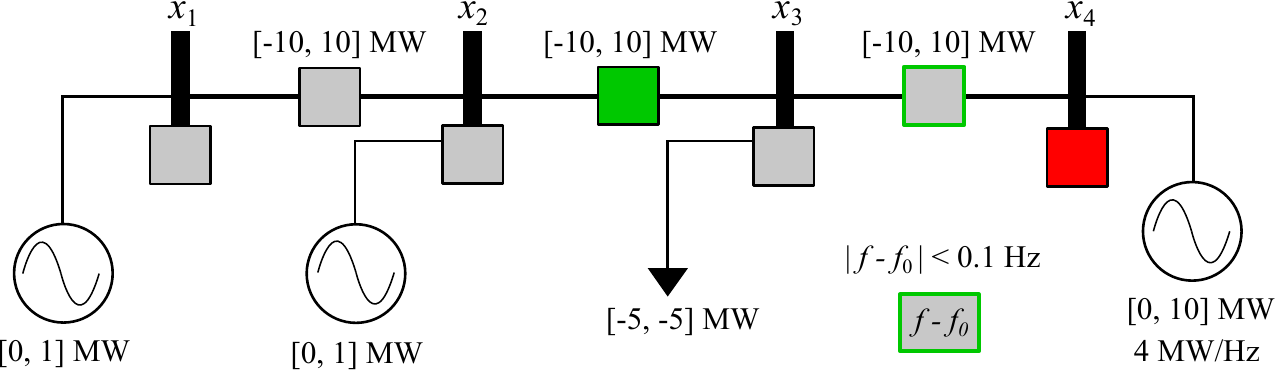} \hspace{3mm}
\includegraphics[width=0.23\columnwidth]{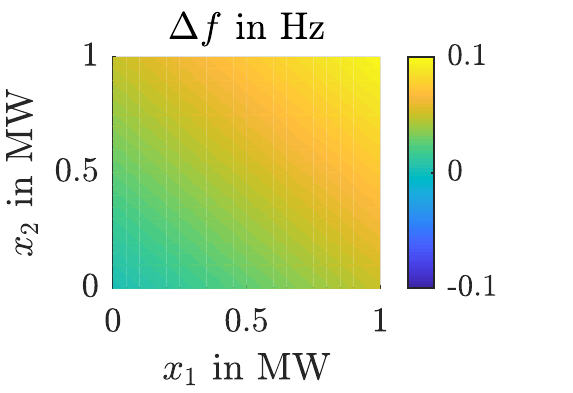}
\subcaption{
$\xc = x_4; \;
\xf =
\begin{bmatrix}
x_1 \! & x_2 \! & x_3
\end{bmatrix}^\suT; \;
\mxS = -0.81; \;
\Mmf =
\begin{bmatrix}
1 \!& 1 \!& 0 
\end{bmatrix}; \;
\vlw = 5.$
}
\vspace{3mm}
\label{subfig:simpleGrid3}
\end{subfigure} \\
%
\begin{subfigure}[h!]{\textwidth}
\centering
\includegraphics[width=0.5\columnwidth]{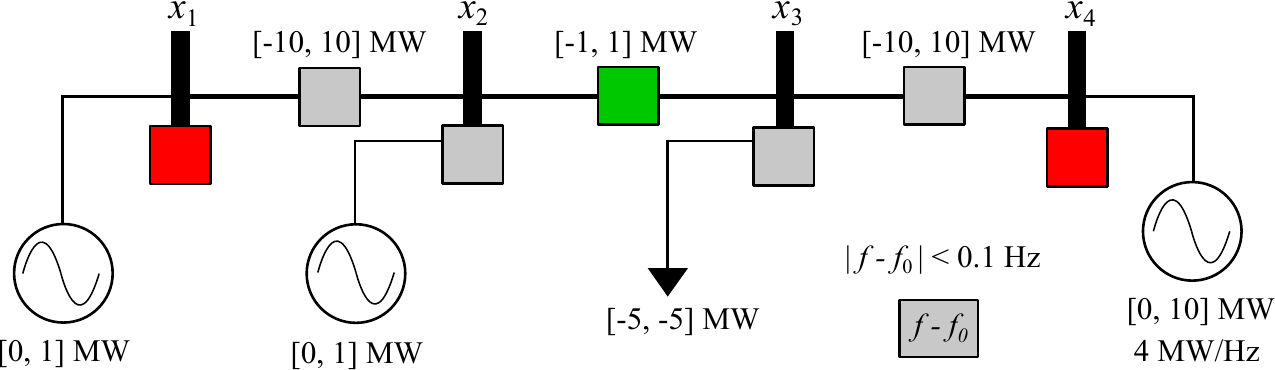} \hspace{3mm}
\includegraphics[width=0.23\columnwidth]{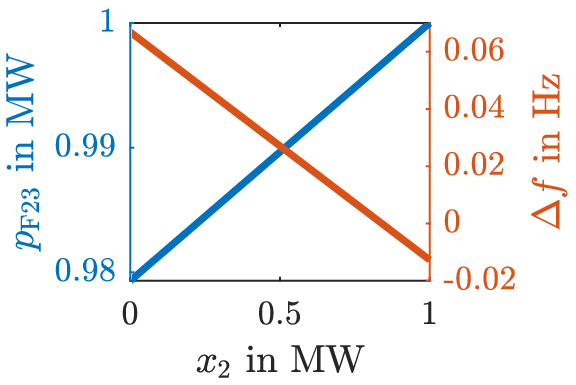}
\subcaption{
$\xc = \begin{bmatrix}
x_1 \\ x_4
\end{bmatrix}; \xf =
\begin{bmatrix}
x_2 \\ x_3
\end{bmatrix};
\Mmf =
\begin{bmatrix}
1 \!& 0
\end{bmatrix};
\mxS = -
\begin{bmatrix}
 0.98 \\
 0.34
\end{bmatrix};
\vlw = 
\begin{bmatrix}
0.98 \\ 4.29
\end{bmatrix}.$
}
\label{subfig:simpleGrid4}
\end{subfigure}

\caption{
Minimal sets of $\mcC$ and $\mcM$ for a simple microgrid.
The gray squares represent potential controller/measurement locations. The selected controllers and measurements are highlighted in red and green, respectively.
Scenarios (a) and (b) have $\mxM=\mxI$, whereas line flows and frequency deviation can also be measured in (c) and (d).
For each scenario, the resulting (non-unique) affine-linear control realization is provided below
together with the behavior of the potentially active constraints for all $\xf\in\Xf$ on the right.
For scenario (c) with multiple, equivalent optimal solutions, the colored frames denote alternative optimal solutions.
The rationale behind the scenarios is as follows:
In (b) the primary control droop is reduced compared to (a).
In (c) we allow for additional measurements.
In (d) we reduce the transfer capacity of the middle link to form an additional active constraint.
}
\label{fig:simpleGrid}%
\end{figure*}

\figref{simpleGrid} shows the considered microgrid consisting of three generators supplying a demand of 5 MW.
It gives the topology of the grid together with the capacity limits of each transmission line and each generator/load.
The generator located at bus 4 provides primary reserve, initially with a droop of 12 MW/Hz and later with 4 MW/Hz.
The maximum allowed frequency deviation is $\pm 0.1$ Hz.
We first assume that all transmission lines have a power transfer capacity of $\pm 10$ MW, which is adequate to avoid grid limitations.
In scenario (d) we add an active line constraint in the middle.

In scenario (a) where only the power set point at each bus may be measured, it is sufficient to control the large generator located at bus 4 for achieving feasible grid operation.
The set points of the remaining smaller generators can be chosen freely and no additional measurement devices are required.

In scenario (b) we reduce the droop of the generator at bus 4 to 4 MW/Hz. 
This makes the measurement of the power injections at buses 1 and 2 necessary. 
Although the power injections at buses 1 and 2 can be chosen arbitrarily, they must be monitored so that the power produced by the generator located at bus 4 can be set appropriately to balance the system within the given frequency tolerance.

In scenarios (a) and (b), where $\mxM = \mxI$, the solutions of the MILP were feasible (and optimal) without further adaptation of $\mcM$ and $\mcC$ and the greedy approach, starting from empty sets $\mcM$ and $\mcC$, produced the same results. 

In scenario (c) the measurement of the line flows and the grid frequency is added to the set of potential measurements, when performing the greedy optimization. 
This allows to reduce the number of measurements to only one.
For this scenario the solution is not unique:
one possibility is to take the measurement of the frequency deviation as controller input, yielding an adapted primary control scheme.
%
An alternative solution that is shown in the figure is to monitor the sum of the outputs of generators 1 and 2 by measuring the line flow between bus 2 and 3 for controlling the set point of the generator at bus 4.
This situation will be very common in future active distribution grids, where individual small scale loads or generators are not able to violate local grid constraints, but their aggregated effect is important to the system.
Since the load is fixed, measuring the line between buses 3 and 4 would be equally informative. 
The feasibility of all these solution candidates was verified via LP \eqref{eq:optTheta}, obtaining valid affine-linear control realizations in all cases.

In scenario (d), we constrain the capacity of the transmission line connecting buses 2 and 3 to the interval $[-1,1]$ MW.
This represents an active grid constraint if the generators at buses 1 and 2 produce at maximum power.
The solution obtained via hill climbing optimization consists of additionally controlling the power injection at bus 1.
Again, several alternative solutions are possible.

In scenarios (a), (b), and (c), the frequency deviation represents an active constraint to the operation of the system.
Observe in \figref{simpleGrid} how in each case the resulting affine-linear control law keeps the frequency deviation inside the feasible region for all values of the non-controlled injections.
In scenario (d), the designed controller also ensures feasible system operation despite the limited power capacity of the middle line.

While for the demonstrated example all solutions can readily be verified manually, 
it shows that the situation may become much more complex in larger grids.
The topological location of generators and loads in the grid is important as well as their capacity and their neighborhood.
An automated algorithm for selecting critical elements to control and/or measure is thus very beneficial for complex networks with distributed generation and transmission lines that are operated close to their technical limits.

In scenarios (c) and (d) the use of the greedy approach is required to deal with $\mxM \not= \mxI$. 
Using the MILP solution as an initial guess for $\mcC$ or starting with empty sets led to the same optimal objective function value.
The solutions for $\mcM$ and $\mcC$ did not always agree exactly, but could be shown to be equally optimal.

The total solver time for all scenarios is shown in Table I.
As expected, the MILP optimization performs faster than the hill climbing optimization for the same instances.
When computing the optimal sets for scenarios (a) and (b), the MILP algorithm was more than 2 times faster than the hill climbing with empty sets.
It was also 1.2 times faster than the hill climbing that uses the MILP solution for $\mcC$ as initial guess,
which corroborates the benefits of such concatenated optimization procedure.

\begin{table}%
\centering
\begin{tabular}{|c|c|c|c|}
\hline
Scenario & MILP  & HC (empty sets) & HC ($\mcC$ from MILP)\\ \hline
(a)      & 77  & 160   & 100  \\
(b)      & 78  & 233  & 135  \\
(c)      & -     & 277  & 171  \\
(d)      & -     & 330  & 176  \\
\hline
\end{tabular}
\caption{Total solver time, in milliseconds, for the proposed optimization algorithms applied to the simple microgrid.}
\label{}
\end{table}

\subsection{IEEE 118 Bus Test Case}
\begin{figure}%
\centering
\begin{subfigure}[b]{\columnwidth}
\includegraphics[width=\columnwidth]{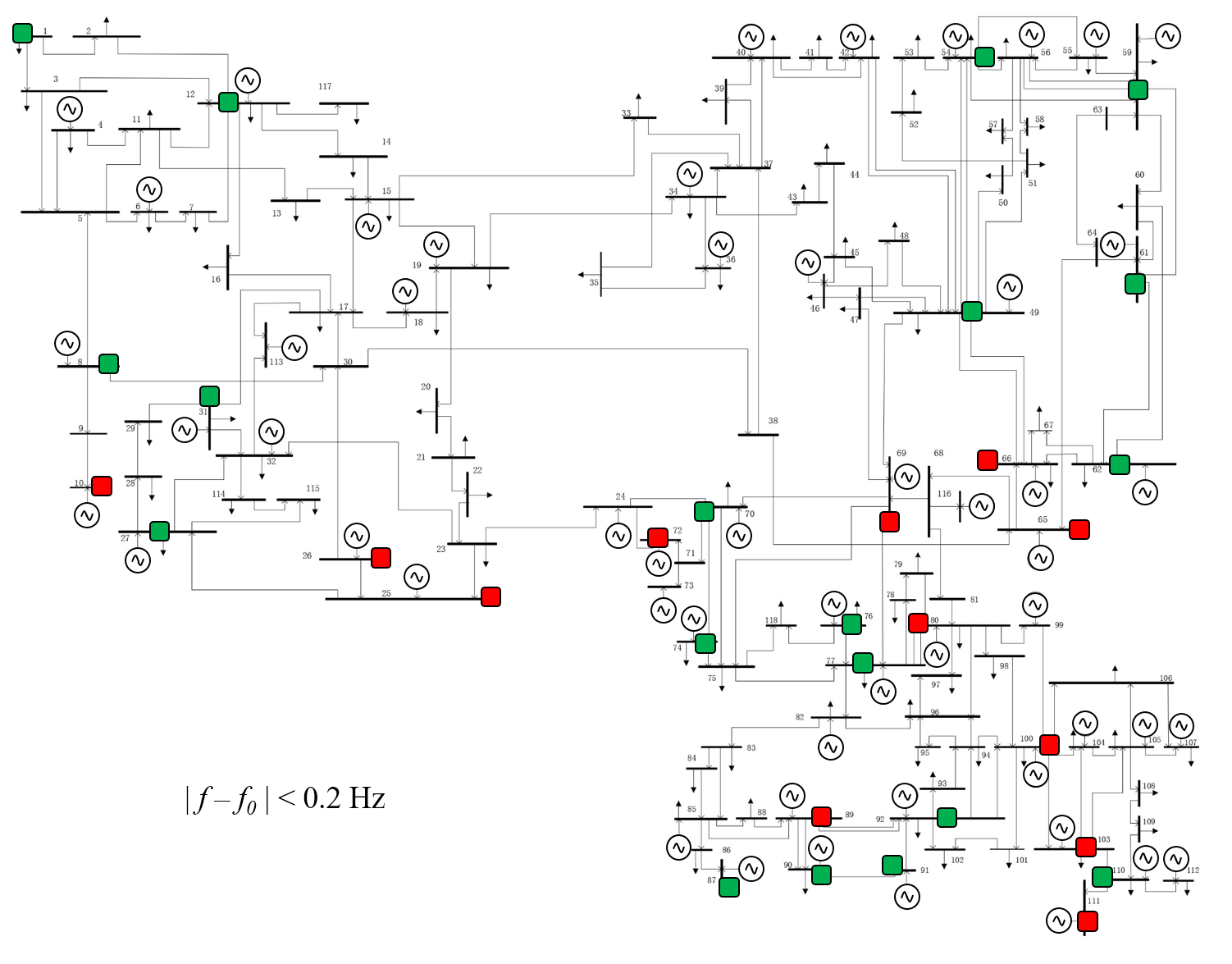}
\subcaption{}
\label{subfig:IEEE118_MILP}
\end{subfigure}
\begin{subfigure}[b]{\columnwidth}
\includegraphics[width=\columnwidth]{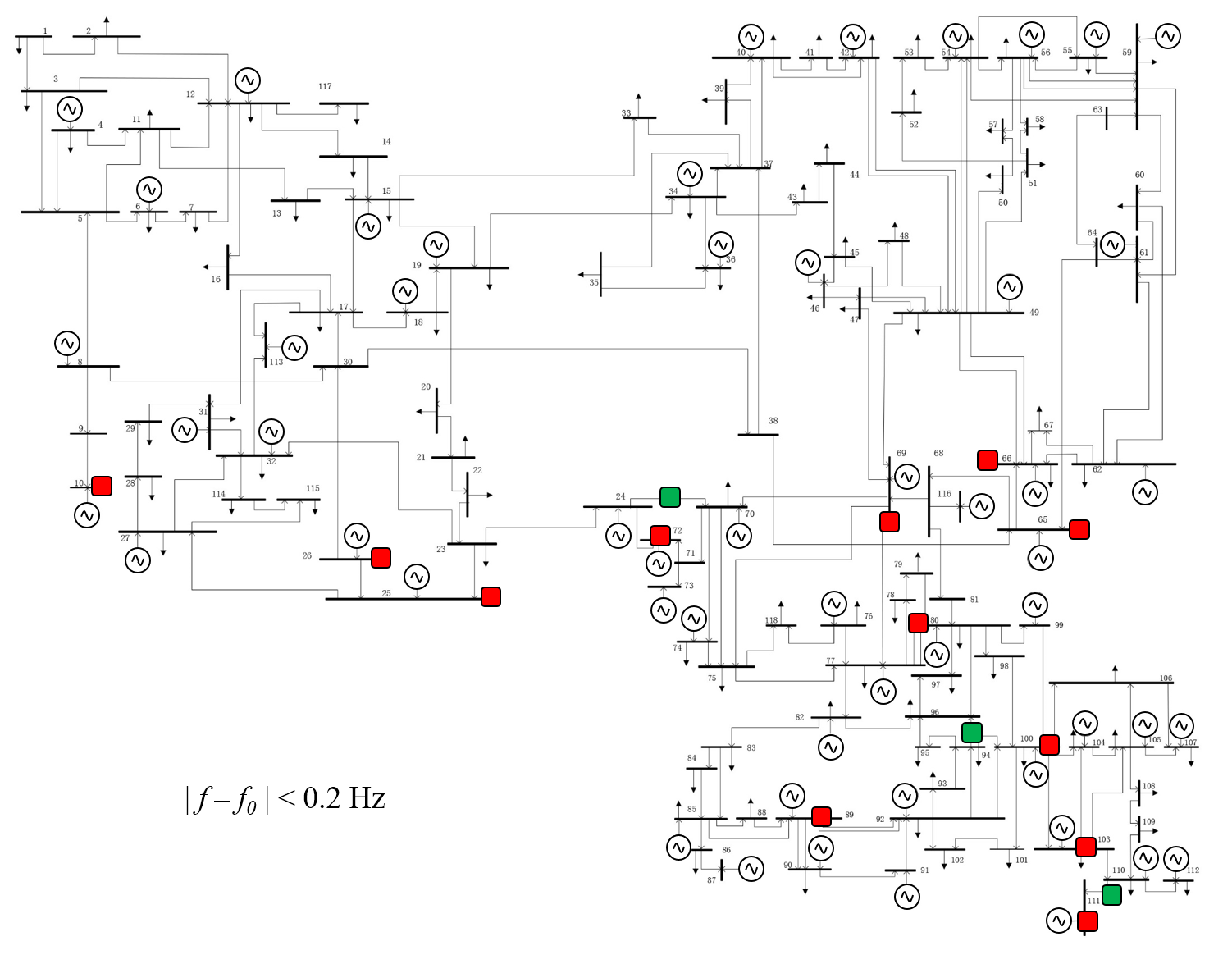}
\subcaption{}
\label{subfig:IEEE118_HC}
\end{subfigure}
\caption{
Minimal sets of controllers and measurements for the modified IEEE 118 bus test case. 
The selected controllers and measurements are highlighted in red and green, respectively. 
(a) Only the nodal power set points may be measured. In this scenario, only 12 controllers and 20 sensors are required to guarantee feasible grid operation. 
(b) The measurement of line power flows and grid frequency deviation are additionally considered as possible. 
In this case only 3 sensors are required.
}
\label{fig:IEEE118Grid}%
\end{figure}

We now analyze the modified version of the IEEE 118 bus test case, see \figref{IEEE118Grid}. 
This power system is composed of 54 generators, 99 loads, and 186 transmission lines. 
The topology of the power system, the load values and the line and generator capacities were taken from \cite{Al-Roomi2015}.
We assume that each generator can be scheduled in the range of 10\%-90\% of its available capacity. 
In addition, we admit 10\% of uncertainty of each load in both directions.
The maximum allowed frequency deviation is taken as $\pm 0.2$ Hz.

We first consider the case when only the power set points may be measured, i.e., $\mxM = \mxI$, see \subfigref{IEEE118_MILP}.
We obtain an optimal set of 12 controller and 20 measurement devices to guarantee feasible grid operation. 
The remaining 96 injections can be left operating free and/or be manipulated deliberately and do not require any monitoring equipment.

To obtain this result, we first use the MILP algorithm and then validate its solution via LP~\eqref{eq:optTheta}.
The obtained $\eta$ is smaller than zero, thereby proving the feasibility and optimality of the MILP solution.
When we initialize the greedy search with empty sets, we obtain a feasible solution consisting of 23 controllers and 9 measurements.
As expected, the obtained solution in this case is larger than the one provided via MILP optimization. 
This confirms that taking the MILP solution as initial guess is beneficial for the greedy search.

We now add the measurements of the line flows and the frequency deviation into the set of potential measurements, see \subfigref{IEEE118_HC}.
This yields in total 305 possible sensor devices. 
We first apply the MILP algorithm and then the greedy one, starting with the controllers identified via the MILP.
As expected, the solution is much sparser than before.
The total number of required sensors is reduced from 20 to 3.
The selected line flows confer a large amount of information
that help avoiding grid capacity violations.

It is insightful to observe the progress of the hill climbing procedure:
buses with major generators connected are selected as controlled nodes first.
The procedure is thus initially reducing the impact of the free variables on the system by controlling the highest uncertain injections first.
When enough controlled nodes were selected, the selection of measurements starts to be significant for the minimization of the cost.
Selected measurements are often related to nodes connected either to large non-controlled generators or to high uncertain loads.
Remaining buses with smaller injections are mostly left unobserved.

Table II shows the obtained solver time for all studied cases.
The solution for case (a) using MILP optimization was found in about 2.57 minutes.
Observe that the solution for case (b) was computed
in about 28 min for the concatenated execution of both algorithms,
compared to the ca. 154 minutes needed by the solver when starting hill climbing with empty sets.
A single verification step using LP~\eqref{eq:optTheta} took less than a second. 

The computation time could further be improved, e.g., by testing not all possible set extensions in each step of the greedy search but using only a representative subset, selected by proximity in the graph. 
Another idea would be to add more than one element in each iteration.
For the control design task described in this paper, however, the achieved computation time seemed acceptable even without these extensions.

\begin{table}%
\centering
\begin{tabular}{|c|c|c|c|}
\hline
Scenario & MILP & HC (empty sets) & HC ($\mcC$ from MILP) \\ \hline
(a) & 2.57    & 13    & 5.78 \\
(b) & -       & 154   & 28 \\
\hline
\end{tabular}
\caption{Total solver time, in minutes, for the proposed optimization algorithms applied to the modified IEEE 118 bus test case.}
\label{}
\end{table}

\section{Outlook}
\label{sec:conclusions}

The theoretic framework and the algorithms developed in this work allow for the efficient identification of critical controllers and measurements in complex power systems with uncertain producers and consumers.
Unlike previous work, we take specific power limitations of lines, generators, and loads into account.
This step strongly improves the applicability in practice,
where our approach will help reducing control costs and efforts and increasing power systems' resilience.

While we have only considered active power in this work, the approach can straightforwardly be applied to linearized power flow models taking into account also reactive power and voltages.
Developing a MILP formulation for condition $\suC_2$ is also possible, but our experiments so far have not yielded satisfying run times. 
\bibliographystyle{IEEEtran}
\bibliography{IEEEabrv,root}

\begin{thebibliography}{10}
\providecommand{\url}[1]{#1}
\csname url@samestyle\endcsname
\providecommand{\newblock}{\relax}
\providecommand{\bibinfo}[2]{#2}
\providecommand{\BIBentrySTDinterwordspacing}{\spaceskip=0pt\relax}
\providecommand{\BIBentryALTinterwordstretchfactor}{4}
\providecommand{\BIBentryALTinterwordspacing}{\spaceskip=\fontdimen2\font plus
\BIBentryALTinterwordstretchfactor\fontdimen3\font minus
  \fontdimen4\font\relax}
\providecommand{\BIBforeignlanguage}[2]{{%
\expandafter\ifx\csname l@#1\endcsname\relax
\typeout{** WARNING: IEEEtran.bst: No hyphenation pattern has been}%
\typeout{** loaded for the language `#1'. Using the pattern for}%
\typeout{** the default language instead.}%
\else
\language=\csname l@#1\endcsname
\fi
#2}}
\providecommand{\BIBdecl}{\relax}
\BIBdecl

\bibitem{Abedi2019}
A.~Abedi, L.~Gaudard, and F.~Romerio, ``Review of major approaches to analyze
  vulnerability in power system,'' \emph{Reliability Engineering {\&} System
  Safety}, vol. 183, pp. 153--172, Mar. 2019.

\bibitem{Yuan2016}
W.~Yuan, J.~Wang, F.~Qiu, C.~Chen, C.~Kang, and B.~Zeng, ``Robust
  optimization-based resilient distribution network planning against natural
  disasters,'' \emph{{IEEE} Transactions on Smart Grid}, vol.~7, no.~6, pp.
  2817--2826, Nov. 2016.

\bibitem{Kalman1960}
R.~E. Kalman, ``A new approach to linear filtering and prediction problems,''
  \emph{Transactions of the ASME--Journal of Basic Engineering}, vol.~82, no.
  Series D, pp. 35--45, 1960.

\bibitem{Li2018}
G.~{Li}, L.~{Deng}, G.~{Xiao}, P.~{Tang}, C.~{Wen}, W.~{Hu}, J.~{Pei},
  L.~{Shi}, and H.~E. {Stanley}, ``Enabling controlling complex networks with
  local topological information,'' \emph{Nature - Scientific Reports}, vol.~8,
  no. 4593, Mar. 2018.

\bibitem{Pequito2016}
S.~Pequito, S.~Kar, and A.~P. Aguiar, ``A framework for structural input/output
  and control configuration selection in large-scale systems,'' \emph{{IEEE}
  Transactions on Automatic Control}, vol.~61, no.~2, Feb. 2016.

\bibitem{Lin1974}
{Ching-Tai Lin}, ``Structural controllability,'' \emph{IEEE Transactions on
  Automatic Control}, vol.~19, no.~3, pp. 201--208, Jun. 1974.

\bibitem{Pasqualetti2015}
F.~{Pasqualetti}, F.~{Dörfler}, and F.~{Bullo}, ``Control-theoretic methods
  for cyberphysical security: Geometric principles for optimal cross-layer
  resilient control systems,'' \emph{IEEE Control Systems Magazine}, vol.~35,
  no.~1, pp. 110--127, Feb. 2015.

\bibitem{Lindmark2018}
G.~Lindmark and C.~Altafini, ``Minimum energy control for complex networks,''
  \emph{Nature - Scientific Reports}, vol.~8, no.~1, Feb. 2018.

\bibitem{Summers2016}
T.~H. {Summers}, F.~L. {Cortesi}, and J.~{Lygeros}, ``On submodularity and
  controllability in complex dynamical networks,'' \emph{IEEE Transactions on
  Control of Network Systems}, vol.~3, no.~1, pp. 91--101, Mar. 2016.

\bibitem{Taha2019}
A.~Taha, N.~Gatsis, T.~Summers, and S.~Nugroho, ``Time-varying sensor and
  actuator selection for uncertain cyber-physical systems,'' \emph{{IEEE}
  Transactions on Control of Network Systems}, vol.~6, no.~2, Jun. 2019.

\bibitem{Molzahn2017}
D.~K. Molzahn, F.~Dörfler, H.~Sandberg, S.~H. Low, S.~Chakrabarti, R.~Baldick,
  and J.~Lavaei, ``A survey of distributed optimization and control algorithms
  for electric power systems,'' \emph{{IEEE} Transactions on Smart Grid},
  vol.~8, no.~6, pp. 2941--2962, Nov. 2017.

\bibitem{Mesanovic2018}
A.~{Mešanović}, U.~{Münz}, and C.~{Ebenbauer}, ``Robust optimal power flow
  for mixed ac/dc transmission systems with volatile renewables,'' \emph{IEEE
  Transactions on Power Systems}, vol.~33, no.~5, pp. 5171--5182, Sep. 2018.

\bibitem{kundur1994power}
P.~Kundur, N.~Balu, and M.~Lauby, \emph{Power System Stability and Control},
  ser. EPRI power system engineering series.\hskip 1em plus 0.5em minus
  0.4em\relax McGraw-Hill Education, 1994.

\bibitem{Lofberg2004}
J.~L{\"{o}}fberg, ``Yalmip : A toolbox for modeling and optimization in
  matlab,'' in \emph{In Proceedings of the CACSD Conference}, Taipei, Taiwan,
  2004.

\bibitem{Al-Roomi2015}
\BIBentryALTinterwordspacing
A.~R. Al-Roomi, ``{Power Flow Test Systems Repository},'' Halifax, Nova Scotia,
  Canada, 2015. [Online]. Available: \url{https://al-roomi.org/power-flow}
\BIBentrySTDinterwordspacing

\end{thebibliography}
%

\end{document}